# Charge-spin current conversion in high quality epitaxial Fe/Pt systems: isotropic spin Hall angle along different in-plane crystalline directions

C. Guillemard[1], S. Petit-Watelot[1*], S. Andrieu[1] and J.-C. Rojas-Sánchez[1*]

[1]Université de Lorraine, CNRS UMR 7198, Institut Jean Lamour, F-54011 Nancy, France

* sebastien.petit@univ-lorraine.fr
* juan-carlos.rojas-sanchez@univ-lorraine.fr

ABSTRACT

We report the growth of MgO[001]//Fe(6 nm)/MgO(7 nm) and MgO[001]//Fe(6 nm)/Pt(6 nm) by molecular beam epitaxy and show that the full characterization by spin-orbit ferromagnetic resonance (SO-FMR) allows the determination of magnetic anisotropies as by classical FMR-only studies. The spin mixing conductance of epitaxial Fe/Pt interface was measured to be $g_{\text{effect}}^{\uparrow\downarrow} = 1.5 \pm 0.5 \times 10^{19} \, \text{m}^2$, and the effective spin Hall angle $\theta_{\text{SHE}}^{\text{effect}}$ was estimated at different in-plane crystalline directions. It was found that $\theta_{\text{SHE}}^{\text{effect}}$ is the same in all directions. When taking into account high enough excitation frequencies to achieve uniform precession of magnetization, the effective spin hall angle for epitaxial Pt in Fe/Pt is $\theta_{\text{SHE}}^{\text{effect}} = 0.051 \pm 0.005$. We address about the proper conditions to determine those relevant spintronics parameters.

TEXT

The conversion of spin current into charge current and vice versa play key roles in new research efforts in spintronics and related applications. This interconversion can be achieved without any external magnetic field or magnetic material in 3-Dimensional systems that exhibit strong spin-orbit coupling [1], namely the spin Hall effect (SHE) [2–4]. The quantification of the efficiency of such interconversion is called the spin Hall angle (SHA). The spin Hall angle determination is thus relevant to find out new materials for applications like in magnetic memories because it will allow reducing power consumption. Large spin Hall angles have been found in heavy metals like Pt [5–8], Ta [9] and W [10,11], and alloys like CuBi [12], AuW and AuTa [13,14]. So far the quantification has been evaluated mostly on sputtered polycrystalline samples. In a heavy metal or alloy layer which is in contact with a layer of a different material a injected spin current might decrease through the interface due to interfacial interactions and the spin Hall angle becomes and effective spin Hall angle $\theta_{\text{SHE}}^{\text{effect}}$ [8,15].

There are several techniques available to evaluate the effective spin Hall angle, or effective spin orbit torque (SOT), like spin pumping ferromagnetic resonance [7,16–19], spin-orbit ferromagnetic resonance [6,20–23], no-local injection in lateral spin valve, current-induced magnetization switching, harmonic measurements, spin Hall magnetoresistance and so on. In all these experiments, in the measurements as well as in the analysis, there are many details and approximations that are often overlooked. In this paper we focus on spin-orbit ferromagnetic resonance (SO-FMR), technique to study epitaxial samples.

**Sample growth**

Epitaxial *s*//Fe(6 nm)/Pt(5 nm) bilayers was grown by molecular beam epitaxy, where *s* stands for the crystalline [001]MgO substrate.
The charge-spin current conversion was evaluated by SO-FMR. For comparison a single Fe layer capped with 7 nm of MgO was grown and characterized by classical FMR method with magnetic field $H_{dc}$ applied in-plane along different crystalline directions.

The Fe layer was grown simultaneously for the *s*//Fe/MgO reference sample as for the *s*//Fe/Pt bilayer as shown in Fig. 1. After the deposition of the Fe layer, half of the sample was covered to deposit 7 nm of MgO, and then the operation is reversed to deposit the Pt layer. In such a way, we have a true Fe reference layer to estimate for instance the effective spin mixing conductance at the Fe/Pt interface. The 2D growth and quality of the sample was monitored in-situ by RHEED as shown in Fig. 1b. The Fe crystalline cell grown rotated by 45° on top of the [001]MgO crystalline cell because of their lattice parameters. That is the direction [110]Fe || [100]MgO and the direction [100]Fe || [110]MgO [24]. In the following we will refer only to the MgO crystalline axes.

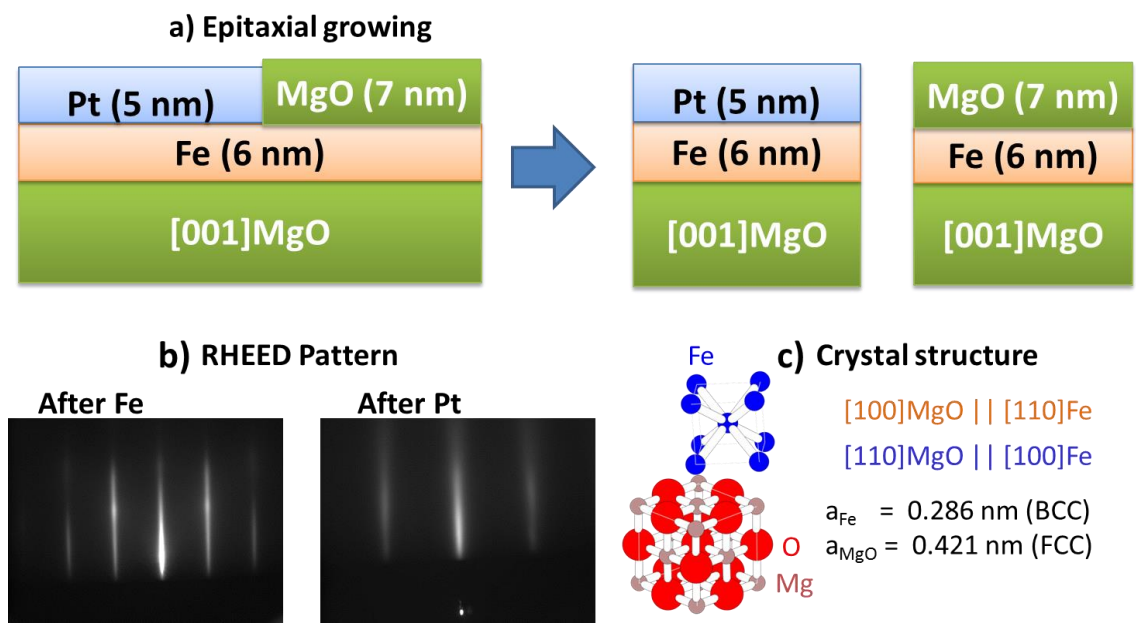

*Fig. 1: (color online)* (a) Schematic of the full stack that has been grown by MBE. After the deposition of the Fe layer, half of the sample was cover to deposit 7 nm of MgO or 5 nm of Pt. b) RHEED pattern shown the good quality and 2D grown of Fe as well as the Pt layers. c) The cubic Fe cell grown rotated by 45° on top of the cubic cell of MgO.

**FMR on epitaxial *s*//Fe/MgO**

Small pieces of about few mm$^2$ were extracted from the center region of *s*//Fe/MgO sample to avoid edge deposition issues and measured by FMR. A grounded coplanar wave guide (GCPW) was used as shown in the Fig. 2a. Hence the radio-frequency magnetic field was maintained transversal (along *y* direction) to the dc magnetic field $H_{dc}$ (along *x*). DC field was applied parallel to the film plane in the present study. The frequency and input power of the microwave source was fixed and $H_{dc}$ was swept around the resonance field as shown in Fig. 2b. The transmitted power is detected using a power diode detector and lock-in technique. The amplitude of the rf power was modulated at $f_{LO}$ (= 433 Hz) and the output signal of the diode is measured at $f_{LO}$. A typical FMR spectrum is shown in the Fig. 2b when $H_{dc}$ is applied parallel to the [100]MgO crystalline axis. The observed symmetrical Lorentz curve verifies that, in this geometry, the absorbed microwave power is proportional to the imaginary part of the magnetic susceptibility $\chi_{yy}$" with practically no losses due to dispersion in the GCPW. We can identify the resonance field $H_{res}$, and the linewidth $\Delta H$ (half width at half maximum). Broadband frequency dependence was studied when $H_{dc}$ is applied along different crystalline directions. The *f*-dependence of the resonance field and linewidth was analyzed and plotted in Fig. 2 (c) and (d), respectively. We can observe in the Fig. 2(c) the typical relationship dispersion *f* vs. *H* in a system with cubic magnetocrystalline anisotropy where the in-plane easy axes are along the [110]MgO directions and the in-plane hard axes are along [100]MgO ones. It is worth to note that every 90 degrees we recover the same relationship without any shift (vertical neither horizontal). This means the lack of additional in-plane uniaxial anisotropy, which is an experimental verification of a high quality sample. After minimizing numerically the free density energy of the system we get the equilibrium position of the magnetization and the resonance condition dispersion following the results of Smit and Beljers [25], and Suhl [26], which formalism has been developed in detail elsewhere [27]. Thus the effective magnetic saturation $M_{eff}$ is 1450 emu/cm$^3$ = 1450 kA/m, and the cubic magnetocrystalline field $\mu_0 H_{cub}$ is -48 mT. We observe that the magnetic damping constant is the same along different crystalline directions with a value of $\alpha_{Fe} = 0.0041 \pm 0.0004$. It is worth to note here that the frequency-independent contribution due to inhomogeneity is quite low, $\mu_0 \Delta H_0 \cong 0.4$ mT, is another verification of high quality epitaxial growth.

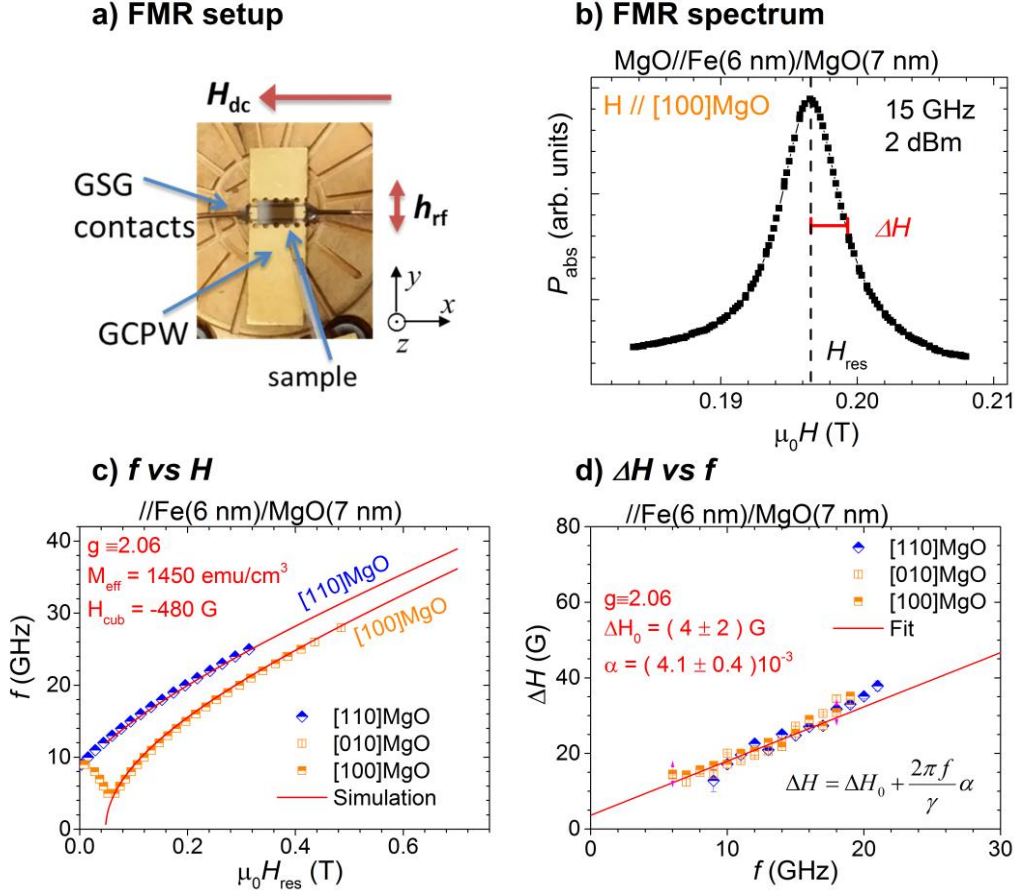

*Figure 2: (a) Picture of the grounded coplanar wave guide GCPW used along with GSG radiofrequency contacts. The sample position, the direction of the dc magnetic field, $H_{dc}$, and the direction of the rf field, $h_{rf}$, in the middle of the signal line are also indicated. (b) Typical FMR spectrum. One can identify the resonance field $H_{res}$ and the linewidth $\Delta H$ for a frequency of 15 GHz when H is applied parallel to the [100]MgO crystalline direction. (c) Dispersion relationship f vs $H_{res}$. We can identify the easy, [110MgO], and in-plane hard axes, [100]MgO, due to cubic magnetocrystalline anisotropy. Lines are numerical simulation as described in the text. (d) Determination of magnetic damping constant $\alpha$.*

## SO-FMR on epitaxial s//Fe/Pt

In order to evaluate the charge-spin current conversion spin-orbit ferromagnetic resonance (SO-FMR), was used. This is also known as spin transfer FMR or spin diode FMR, which is somehow the reciprocal dynamic effect of spin pumping FMR. We inject directly the microwave frequency charge current in the s//Fe/Pt slabs which is converted into spin current inside the Pt layer due to the SHE. Therefore an oscillating spin current is injected from Pt into Fe layer inducing precession of its magnetization. This, in turn, leads to an oscillatory radiofrequency resistance which mixed with the rf current allows, at the resonance condition, dc voltage detection across the slab using a bias tee [6,20–23]. The dc voltage is picked up at 45° of the applied $H_{dc}$ (see inset in Figure 3). The dc voltage is composed of a mix between a symmetrical Lorentzian function and an antisymmetrical one around the resonance field $H_{res}$. The amplitude of each contribution is $V_{sym}$ and $V_{anti}$. We use the following general function to fit the mixed voltage measured considering also an offset V$_{offset}$:

$$V_{mix} = V_{offset} + V_{sym}\frac{\Delta H^2}{\Delta H^2 + (H-H_{res})^2} + V_{anti}\frac{(H-H_{res})\Delta H}{\Delta H^2 + (H-H_{res})^2} \quad (1)$$

The slabs for SO-FMR were patterned by standard UV lithography and have lateral sizes of 20 μm × 90 μm. Ti/Au electrodes were deposited by lift off technique. In the inset of figure 3(a) is shown a picture of one device along with grounded-signal-grounded, GSG, rf contacts. Figure 3 shows raw data of the dc voltage measured when $H_{dc}$ is applied along different crystalline axes. After the analysis of this broadband frequency dependence results we show that by SO-FMR we can also account for the in-plane magnetic anisotropies, i.e, the easy (hard) axes when $H_{dc}$ is parallel to [110]MgO ([100]MgO). That is summarized in figure 4(a) where we can identify similar results of only FMR as observed in Fig. 2(c). This means that despite the pick-up voltage is at 45° of the applied field in the SO-FMR experiment, what matters is the direction along which is applied $H_{dc}$ with respect to the crystalline axis of the samples. So far in our knowledge, this is the first time that it is shown such results in a system with cubic anisotropy. The magnetic damping in Fe/Pt results $\alpha_{Fe/Pt} = 0.0065 \pm 0.0004$ as shown in figure 4(b). With the results from the Fe reference layer ($\alpha_{Fe} = 0.0041 \pm 0.0004$) we can calculate the effective spin mixing conductance $g_{eff}^{\uparrow\downarrow}$ following standard spin pumping theory [8,13,28]:

$$g_{eff}^{\uparrow\downarrow} = \frac{4\pi M_{eff} t_{Fe}}{g\mu_B}(\alpha_{Fe/Pt} - \alpha_{Fe}) \quad (2)$$

$\alpha_{Fe}$ is quite large. To compute $g_{eff}^{\uparrow\downarrow}$ we use the difference of the magnetic damping ($\alpha_{Fe/Pt} - \alpha_{Fe}$) which gives the contribution due to spin pumping only since the Fe layer is a true reference sample as described above.

This results in $g_{effect}^{\uparrow\downarrow} = 1.5 \pm 0.5 \times 10^{19}$ which is lower than previously reported for FM/Pt systems, even in epitaxial Fe/Pt samples studied by spin pumping [29,30]. Critical for this result is that the Fe layer is exactly the same for both reference Fe/MgO and Fe/Pt bilayer which is crucial especially in epitaxial samples where damping is much more sensitive to any defect (which is not the case in polycrystalline samples). Future work will include measuring the reference damping with the same SO-FMR method using a structure like Fe/Cu/MgO and thickness dependence.

In the simplest model, the quantification of the effective spin Hall angle $\theta_{SHE}^{effect}$ is proportional to the ratio of the symmetrical Lorentzian voltage over the anti-Lorentzian one, V$_{symm}$/V$_{anti}$. When the resonance field is large enough to consider uniform precession of the magnetization we can use the simplest model to calculate the effective spin Hall angle [6,21–23]:

$$\text{If} \quad H_{sat} < H_{res} \rightarrow \quad \theta_{SHE}^{eff} = \frac{J_s}{J_c} \cong \frac{V_{sym}}{V_{anti}}\frac{e\mu_0 M_s t_{Fe} t_{Pt}}{\hbar}\left[1 + \frac{4\pi M_{eff}}{H_{res}}\right]^{1/2} \quad (3)$$

where we have highlighted the condition of uniform precession of the magnetization ($H_{res} > H_{sat}$), with $H_{sat}$ being the saturation field. In the in-plane hard axes, the field necessary to get the saturation state is above 0.1 T, obtained after M(H) cycles in PPMS-VSM (not shown). The saturation magnetization is $M_s$ = 1675 emu/cm$^3$, close to the bulk value (1714 emu/cm$^3$) [31]. We can calculate the four-fold magnetocrystalline anisotropy ($K_{cub} = H_{cub} \times M_s / 2$). It results in our Fe film that $K_{cub} \approx 3.9 \times 10^5$ erg/cm$^3$, below the bulk value of Fe ($\approx 4.8 \times 10^5$ erg/cm$^3$) [31]. As shown in Figure 4(c) and using eq. (3), we find that for Pt in Fe/Pt system $\theta_{SHE}^{effect} = 0.051 \pm 0.008$, which is similar that previous effectives values reported [6,8]. Moreover, within the experimental resolution, we have found similar values when H is applied along different crystalline directions once it is reached the condition: $H_{res} > H_{sat}$ as displayed in figure 4(c).

Additionally, we have grown s//Fe(6)/Pt(t) with t between 1 and 10 nm, and then patterned in double Hall bar to measure the electrical resistance. From the Pt-thickness dependence of such results we have estimated the resistivity of Pt layer whose value is $\rho_{Pt} = 15.6 \pm 0.5$ μΩ.cm along different in-plane crystalline directions. It has been shown that Elliot-Yafet mechanism dominates the conduction electron spin relaxation in Pt [15,32,33] where the product $\rho_{Pt} \times l_{sf}$ is a constant. Experimental values of such a product [8,32] are close to the theoretical one, $\rho_{Pt} \times l_{sf} \approx 0.61\ f\Omega m^2$ [15]. Thus we can consider that in this epitaxial system the spin diffusion length of Pt is $l_{sf} \approx 3.9$ nm. The estimated product $\theta_{SHE}^{effect} \times lsf \approx 0.19$ nm results close to the value reported in most studies where Pt-thickness dependence was performed [8].

Some other systems that showed isotropic results despite in-plane anisotropies can be found in the case of in-plane exchange bias and spin-orbit torque in AFM/FM [34], or isotropic damping in epitaxial Fe [35], for similar level of thicknesses. However it just has been shown that for ultrathin Fe layer (0.8 nm) we can detect anisotropic magnetic damping [35], as predicted theoretically for epitaxial magnetic layers [36]. Ultrathin magnetic layers are out of the scope of the present study but it is a very interesting perspective to enhance this study especially with epitaxial Heusler alloys showing ultralow damping [37].

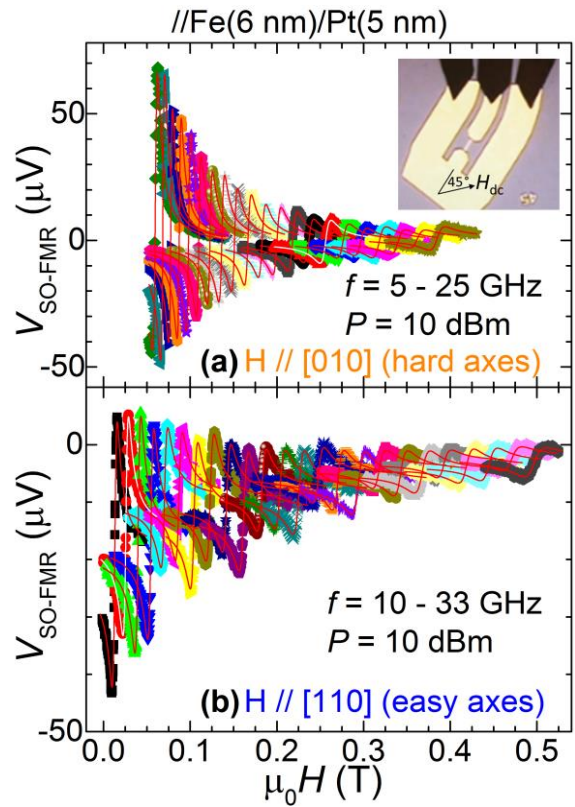

Figure 3. Raw data of the SO-FMR scans. (a) $H_{dc}$ is applied parallel to the [010]MgO crystalline direction (scans shown only for f ≥ 6 GHz). The inset is a picture of a device along with Ti/Au electrodes and GSG rf contacts. (b) $H_{dc}$ is applied parallel to the [110]MgO crystalline direction. $H_{dc}$ is always applied at 45° of the slab.

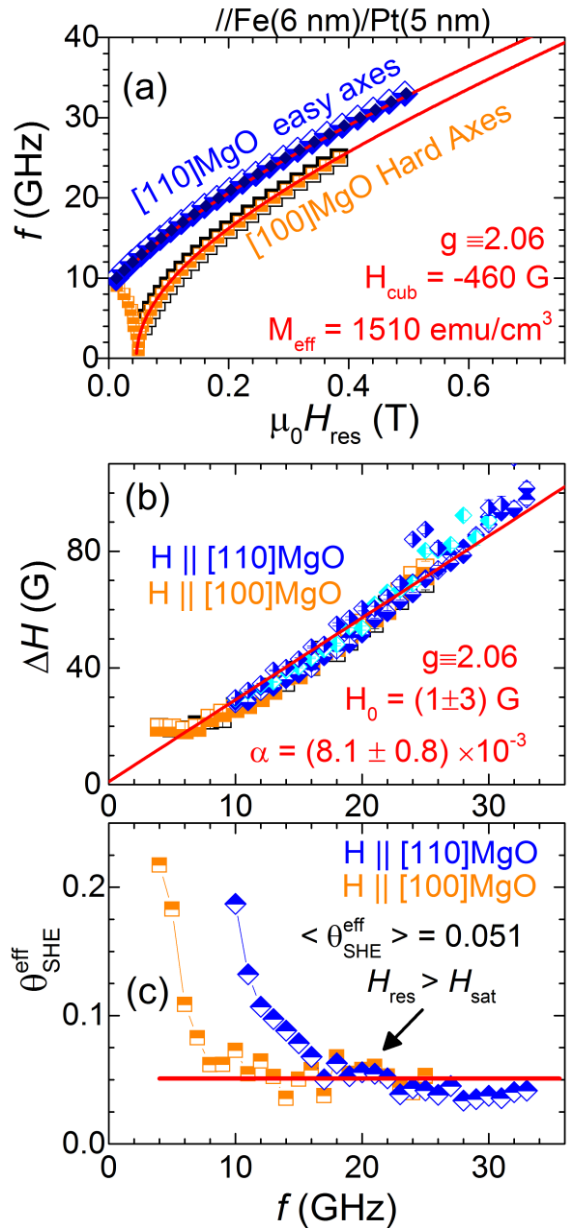

Figure 4. Results of the SO-FMR study. (a) Dispersion relationship and magnetic anisotropies determination. (b) Linewidth and damping constant determination. (c) Determination of effective spin Hall angle. We observe that for f > 10 GHz (>15 GHz) along the hard (easy) axes it reaches a constant value (as expected it does not depend on the frequency). The resonance field for such a frequencies are above 0.1 T, *i.e.*, well above saturation field so uniform precession of magnetization is reached.

In summary, we have performed FMR and SO-FMR in thin highly epitaxial, Fe and Fe/Pt samples respectively. From both methods we show that one can access detailed magnetic anisotropies. Particularly we have accounted for the cubic magnetocrystalline anisotropy of Fe. Using exactly the same Fe bottom layer, a low value of spin mixing conductance in Fe/Pt was measured. We show that considering uniform precession of the magnetization (resonance field above saturation field), the effective spin Hall angle can be determined in a reliable way for FM/Pt systems, and, it is independent of the microwave frequency. Furthermore, we show that the magnetic Gilbert damping constant as well as the effective spin Hall angle are isotropic in epitaxial system for the level of thickness in the present study, Fe(6 nm)/Pt(5 nm).

Our results also highlight the importance of taking care of the proper conditions to the estimation of the effective charge-spin current conversion efficiency or spin Hall angle, and the effective spin mixing conductance in epitaxial FM/HM systems. Those parameters and their proper determination are relevant for spintronics applications.

ACKNOWLEDGMENTS

This work was supported partly by the french PIA project "Lorraine Université d'Excellence", reference ANR-15-IDEX-04-LUE. By the ANR-NSF Project, ANR-13-IS04-0008- 01. Experiments were performed using equipment from the TUBE—Daum funded by FEDER (EU), ANR, the Region Lorraine and Grand Nancy. We thank L. E. Ocola for a proof reading of the manuscript.